\documentclass[manuscript]{aastex}

\shorttitle{Magnetic Field Strength in Solar Corona}
\shortauthors{Kim et al.}

\begin{document}


\title{Magnetic Field Strength in the Upper Solar Corona Using White-light Shock Structures Surrounding Coronal Mass Ejections}


\author{R.-S. Kim,\altaffilmark{1,2} N. Gopalswamy,\altaffilmark{1} K.-S. Cho,\altaffilmark{1,2,3} and S. Yashiro\altaffilmark{1,2}}
\affil{NASA Goddard Space Flight Center, Greenbelt, MD 20771}
\email{rok-soon.kim@nasa.gov}

\and

\author{Y.-J. Moon\altaffilmark{4}}
\affil{School of Space Research, Kyung Hee University, Yongin-shi, 446-701, Korea}


\altaffiltext{1}{NASA Goddard Space Flight Center, Greenbelt, Maryland, USA.}
\altaffiltext{2}{Department of Physics, The Catholic University of America, Washington, D. C., USA.}
\altaffiltext{3}{Korea Astronomy and Space Science Institute, Daejeon, Korea}
\altaffiltext{4}{School of Space Research, Kyung Hee University, Yongin-shi, 446-701, Korea}


\begin{abstract}

To measure the magnetic field strength in the solar corona, we examined 10 fast ($\geq 1000$ km s$^{-1}$) limb CMEs which show clear shock structures in SOHO/LASCO images.
By applying piston-shock relationship to the observed CME's standoff distance and electron density compression ratio, we estimated the Mach number, Alfven speed, and magnetic field strength in the height range 3 to 15 solar radii ($R_s$).
Main results from this study are: (1) the standoff distance observed in solar corona is consistent with those from a magnetohydrodynamic (MHD) model and near-Earth observations; (2) the Mach number as a shock strength is in the range 1.49 to 3.43 from the standoff distance ratio, but when we use the density compression ratio, the Mach number is in the range 1.47 to 1.90, implying that the measured density compression ratio is likely to be underestimated due to observational limits; (3) the Alfven speed ranges from 259 to 982 km s$^{-1}$ and the magnetic field strength is in the range 6 to $105 mG$ when the standoff distance is used; (4) if we multiply the density compression ratio by a factor of 2, the Alfven speeds and the magnetic field strengths are consistent in both methods; (5) the magnetic field strengths derived from the shock parameters are similar to those of empirical models and previous estimates.
\end{abstract}



\section{Introduction}

The solar corona is the plasma atmosphere of the Sun, extending millions of kilometers into space. One of the major issues of the solar corona is to measure the magnetic field, which determines the coronal structure and dynamics from the upper chromosphere out into the heliosphere. Evidence for magnetic field in the corona has been found in several kinds of observations, but only a few of them give magnetic field information since the coronal plasma is optically thin (Wiegelmann 2008). These observations are possible only in a limited spatial extent and the magnetic field has to be derived with some uncertainties in interpretation.

One of the main techniques to estimate the coronal magnetic field is the optical observations of vector magnetic fields in the photosphere and their extrapolation into the corona. Lin et al. (2000) presented the magnetic field strength in the inner corona based on Stokes V circular polarization profiles. Solanki et al. (2003) reported three-dimensional magnetic field topology in an interaction region near the base of the solar corona from the measurement of Stokes vector. Their results of magnetic field strength, $B$ are from tens to several hundreds $G$ in a very limited region ($r < 0.5 R_s$). The extrapolations of the photospheric magnetic field into the solar corona depend on the assumptions such as low $\beta$ plasma, which may not be valid in the outer corona (Gary 2001). Radio data may also be used as a diagnostic of coronal magnetic structure. Lee et al. (1999) used radio observations of an active region to examine the coronal magnetic field obtained via a nonlinear force-free field extrapolation of a photospheric vector magnetogram. Ramesh et al. (2010) presented that the estimated values of $B$ at two different distances at 1.5  and $1.7 R_s$ from the observations of circularly polarized thermal radio emission from solar coronal streamers are 6 $\pm 2 G$ and 5 $\pm 1 G$, respectively. It should be noted that the above techniques can be applied to measure the magnetic fields only in the inner corona ($r < 2 R_s$).

Several studies have been performed to measure magnetic fields in the corona from the band splittings in type II radio bursts (Smerd et al. 1974). Vrsnak et al. (2004) measured the interplanetary magnetic field strength using the band splitting of coronal and interplanetary type II bursts. But, they considered only four bins in the distance range of 25 to 225 $R_s$ due to difficulties in the observation, such as a low signal-to-noise ratio and small amount of data. Cho et al. (2007) used band splitting of coronal type II radio bursts and obtained coronal magnetic field strength of 1.3 to 0.4 $G$ in the height range of 1.5 to 2 $R_s$. Faraday rotation techniques have been occasionally used in estimating the magnetic field strengths at several solar radii (Patzold et al. 1987; Spangler 2005; Ingleby et al. 2007).

It is known that the observations in solar corona approximately follow the empirical formulas $B(r)=0.5 (r/R_{s} -1)^{-1.5} G$ for active regions (Dulk $\&$ McLean 1978), and $B(r)=2.2 (r/R_{s})^{-2} G$ for quiet regions (Mann et al. 1999). Since these formulas were made from the observations of lower corona below $\sim 10 R_{s}$, we may wonder whether  these radial dependencies of $B$ are still effective in the upper coronal region up to several tens of solar radii. The observations by Large Angle and Spectrometric COronagraph (LASCO) on aboard the SOlar and Heliospheric Observatory satellite (SOHO) (Brueckner et al. 1995) enable us to study the upper corona in the range $1.5 R_s < r < 30 R_s$.

These observations have shown that the speeds of coronal mass ejections (CMEs) vary from a few hundred to more than 2500 km s$^{-1}$ (Yashiro et al. 2004; Gopalswamy 2010). CMEs with speeds in excess of the ambient Alfven speed, which is characterized by the magnetic field and plasma density, may drive fast mode MHD shocks. As a shock signature associated with CMEs, streamer deflections have been suggested and observed (Gosling et al. 1974; Michels et al. 1984; Sheeley et al. 2000). Recently, several studies have shown that the CME-driven shocks could be directly observed in white-light coronagraph images, under suitable conditions from the analysis of whith light (Vourlidas et al. 2003; Gopalswamy et al. 2009a; Ontiveros \& Vourlidas 2009; Bemporad \& Mancuso 2010; Gopalswamy 2010) and EUV (Gopalswamy et al. 2011; Ma et al. 2011; Kozarev et al. 2011) images as well as from magnetohydrodynamic (MHD) simulations (Stepanova \& Kosovichev 2000;  Manchester et al. 2004). If these structures are truly the shocks, there should be noticeable rapid rises in pressure, temperature and density of the flow and applicable to the piston-shock relationship (Eselevich \& Eselevich 2010).

In piston-driven shocks such as the Earth's bow shock, there are several parameters that can govern the shock shapes: the size and shape of the obstacle, the electron density compression ratio, the upstream Mach number, and the standoff distance (the distance between the obstacle and its shock nose). Russell and Mulligan (2002) applied the relation between shock standoff distance to a CME near-Earth to explain the curvature of the driving interplanetary CME (ICME). Gopalswamy \& Yashiro (2011) measured the standoff distance of a CME-driven shock in the corona using SOHO/LASCO and the Solar TErrestrial RElations Observatory (STEREO) images and derived the coronal magnetic field in the range of 6 to 23 $R_s$.

In this paper, we consider a large number of shock-driving CMEs identified in the SOHO/LASCO images and use the standoff distance technique (Gopalswamy \& Yashiro 2011) for measuring the coronal magnetic field. Following the case study of Gopalswamy \& Yashiro (2011), we estimate the magnetic field radial distributions in the upper corona using the standoff distance technique. We also use the density compression ratio across the shock to determine the magnetic field and compare the results with those from the standoff distance technique. In addition, we compare the magnetic field distributions with those in previous studies.  We also examine the physical properties of the upstream medium from different techniques for consistency. The paper is organized as follows. The data and methodology are described in Section 2. In section 3, we present the estimations of the coronal Alfven speeds and magnetic field strengths using the standoff distance and the density compression ratio methods. A brief summary and conclusions are given in Section 4.

\section{Data and Methodology}

\subsection{Data Selection}

For the analysis of CME-driven shocks, we selected good candidates
which show clear signatures of discontinuity ahead of the CMEs as
they propagate from the Sun. The left panel of Figure 1 shows an
example of the shock structure observed by the SOHO/LASCO
coronagraph. The shock structure appears as a diffuse feature
surrounding the CME as indicated by an arrow. Assuming that the
leading edge of the diffuse structure is the piston-driven shock, we
measure the shock parameters such as the standoff distance and
electron density compression ratio as indicators of the shock
strength.

To select a sample of CME-driven shock structures, we used the
following procedure: (1) we selected fast CMEs ($\ge 1000$ km
s$^{-1}$)  from 1996 to 2007 using the SOHO/LASCO online CME
catalog\footnote{http://cdaw.gsfc.nasa.gov/CME$\_$list/index.html}
(Yashiro et al. 2004; Gopalswamy et al. 2009b), since these CMEs are
fast enough to drive shocks (see Gopalswamy et al. 2008a); (2) we
checked only CMEs associated with M and X-class solar flares whose
source locations are  close to the limb ($> 60^{\circ}$) to minimize
projection effects; (3) we used the events which show clear shock
structures in at least three frames within C2 and/or C3
fields-of-view. Although we identified 104 CMEs with shock
structures, many were too faint to measure the standoff distance or
they have only one or two frames which show shock signatures. We
also excluded CMEs, which had preceding CMEs within 12 hours, since
the pre-events could significantly disturb the ambient conditions
including the upstream density and the Alfven speed (Eselevich \&
Eselevich 2011). Finally we selected only 26 frames corresponding to
10 events that show relatively clear shock features. These events
mainly occurred during the solar maximum phase of solar cycle 23.
Table 1 summarizes the basic information of these events. We also
list the occurrence of metric and/or DH Type II radio bursts in the
third column since a shock in the leading edge of the CME could be
the source of a type II radio burst (Gopalswamy et al. 2005; Cho et
al. 2011).

\subsection{Standoff Distance Ratio}

The standoff distance, $\Delta R$, in the CME-driven shock structure
is defined as the distance from the front of a CME to its shock nose
in the radial direction as shown in the right panel of Figure 1. The
standoff distance of a strong shock is shorter than that of a weak
shock when we consider the same CME size. Since the standoff
distance is proportional to the size of CME (Russell \& Mulligan
2002; Manchester et al. 2004), we measured the curvature radius of
the CME, $R_{c}$ and determined the ratio of $\Delta R$ to $R_{c}$
as an indicator of the shock strength. $\Delta R$ and $R_{c}$ can be
determined directly from the coronagraph images. The measurement of
the standoff distance ratio, $\Delta R/R_{c}$ is made as follows:
(1) to determine $R_{c}$, we fitted a circle to the CME front in the
SOHO/LASCO running difference image (see the blue circle in the
right panel of Figure 1); (2) we then measured the distance from the
CME front and the leading edge of the diffuse structure in the
radial direction as $\Delta R$ (red straight line); (3) we
considered the position of the shock nose as the shock height. The
central position angle (PA) of the CMEs, the shock height, and
$\Delta R/R_{c}$ are listed in the fourth, fifth, and sixth columns
of Table 1, respectively for the 10 events.

Figure 2 shows the variation of $\Delta R/R_{c}$ with heliocentric
distance for the 10 events, roughly scattered in the range 0.19 to
0.78 (mean=0.34) at the height range from 3.1 to 15.3 $R_{s}$. We
included the standoff distance ratios for a single event from
Gopalswamy \& Yashiro (2011) indicated by red circles. For
comparison, we have also plotted the standoff distances of 7
magnetic clouds (MCs) observed at near-Earth interplanetary (IP)
space, which are selected from the MC list of Gopalswamy et al.
(2008b) satisfying the condition $ 10^\circ E <$ Longitude $
<10^\circ W$ to confirm they pass the Earth by their noses. We
calculated the standoff distances using the time difference of IP
shock and sheath, and MC's speed information from the list and we
assumed the radius of MC's curvature is 0.4 AU as suggested by
Russell \& Mulligan (2002). As a result, the mean of the standoff
distance ratio for 7 MCs is 0.33, which is similar to that of
near-Sun shocks (see Maloney \& Gallagher 2011). The standoff
distances are the same (0.22) for two events, so the data points
overlapped in Figure 2.

We also compared our result with the standoff distance ratio of Manchester et al. (2004) who presented a three-dimensional numerical ideal MHD model describing the time-dependent expulsion of a CME. According to their simulation, the standoff distance of the shock is $4.3 R_{s}$ when the CME's front is at $40 R_s$. Then the shock front reaches 1 AU ($\sim 215 R_s$), $16 R_s$ ahead of the CME. At the two distances of $44.3 R_s$ and $215 R_s$, the ratios of standoff distance are 0.27 and 0.19, respectively, when we take $R_{c}=0.4r$ (Russell \& and Mulligan 2002). We also added their results to Figure 2. As shown in the figure, our $\Delta R/R_{c}$ values are comparable to those from the numerical CME model and near-Earth observations.

\subsection{Electron Density Compression Ratio}

One of the quantities needed in the estimation of the coronal magnetic field is the upstream electron density, which can be estimated from the inversion of polarized brightness (pB) measurements (Gopalswamy \& Yashiro 2011). Van de Hulst (1950) derived a parametric representation for the electron density, $\rho$, as a function of the radial distance from the Sun. This method has been widely applied to obtain radial profiles of the coronal electron density from calibrated white-light images. Hayes et al. (2001) extended the Van de Hulst method to take advantage of the extensive LASCO archive of total brightness images. Polarized brightness images are obtained only twice a day, while total brightness images are obtained with much higher cadence.

To measure the downstream/upstream electron density compression ratio, we adopted the method of Hayes et al. (2001) to LASCO C2 and C3 total brightness (tB) images instead of pB images since it is very hard to obtain the density compression ratio by using the pB images, which have very poor time cadence (2 to 3 frames/day). The detailed procedure to obtain density compression ratio is as follows: (1) for each frame, we selected LASCO/C2 and C3 images, which are in the time window starting 4 hours before the associated eruption and ending 4 hours after the last CME observation; (2) we plotted the radial profile of the electron density at the position angle (PA) corresponding to the nose of the shock; (3) we measured the electron density at the shock height in the radial profile.

Figure 3 shows the temporal variation of the density for a given
shock height for the 2001 April 1 event. This figure shows the
density jump after the CME's first appearance indicated by the
arrow. We calculated the downstream/upstream electron density
compression ratio by dividing the maximum electron density,
$\rho_{d}$ by the average of upstream electron densities, $\rho_{u}$
as marked by the solid lines. We assumed a nominal depth of 1 $R_s$
for all the events because it is a convenient scale and is likely a
good upper limit (Ontiveros \& Vourlidas 2009). The density
compression ratios for the 10 events are in the range of 1.00 and
1.91 (mean=1.18) as shown in Figure 4 and Table 1. In the inner
region below $5 R_{s}$ the compression ratio is relatively higher
than in the outer region, but it is still low and the ratio has a
trend to be close to 1 as the heliocentric distance increases as
indicated by the polynomial fitting result (dotted line). This
result shows different tendency from the Figure 8 of Eselevich \&
Eselevich (2011), which shows that the density compression ratio
increases with the distance. We speculate that the difference is
from the shock size $l$ along the line-of-sight. They assumed $l$ as
6.5 $R_s$ while we used 1 $R_s$. Note that their average shock
height (18.6 $R_s$) is substantially higher than ours (8.2 $R_s$).

\subsection{Shock Speed}

We determined the shock speeds, $V_{SH}$, at 26 shock positions by
subtracting the ambient solar wind speed from the upstream shock
speed, which is measured from two successive frames. The
distribution of the shock speed is shown in Figure 5. The shock
speed ranges from 705 km s$^{-1}$ to 2132 km s$^{-1}$ (mean=1288 km
s$^{-1}$).  The solar wind speed profile was taken from the
empirical relation obtained by Sheeley et al (1997). To compare with
Alfven speed, we added the Alfven speed profile obtained using the
models of magnetic field and plasma density (Dulk $\&$ McLean 1978;
LeBlanc et al. 1998; Mann et at. 1999; Gopalswamy et al. 2001;
Eselevich \& Eselevich 2008). As shown in the figure, all events
have speeds faster than the Alfven speeds and hence can form shocks.
We have also listed the shock speeds, $V_{SH}$,  in the eighth
column of Table 1.

\subsection{Shock Mach Number}

It is well known that the density compression ratio is related to the compressibility of the medium and the upstream Mach number (Landau \& Lifshitz 1959). According to a modified method suggested by Farris \& Russell (1994) for low Mach numbers (weak shock), the density compression ratio is expressed by
\begin{equation}
{\rho_{d} \over \rho_{u}} = {{(\gamma+1)(M^{2}-1)} \over {(\gamma-1) M^{2} +2}},
\end{equation}
where $\gamma$ is the ratio of specific heats and $M$ is the upstream Mach number. Seiff (1962) empirically showed that the standoff distance of a bow shock, which is normalized by the radius of the obstacle, is  linearly proportional to the inverse density ratio. Then Farris \& Russell (1994) modified this relationship to consider the radius of curvature ($R_c$) of the obstacle and the standoff distance ratio can be given by
\begin{equation}
{\Delta R \over R_{c}}=0.8{\rho_{u} \over \rho_{d}}.
\end{equation}
This yields a relationship between the standoff distance ratio and the Mach number:
\begin{equation}
{\Delta R \over R_{c}}=0.8 {{(\gamma-1) M^{2}+2} \over {(\gamma+1) (M^{2}-1)}},
\end{equation}
which indicates that for a weak shock, as M increases, the standoff distance ratio decreases.

If we measure the standoff distance ratio and the density compression ratio, we can calculate the upstream Mach number by rewriting the equation (3) and (1) as
\begin{equation}
M_{\Delta R}^{2} = {{\Delta R/R_{c} (\gamma+1) + 1.6} \over {\Delta R/R_{c} (\gamma+1)-0.8(\gamma-1)}},
\end{equation}
and
\begin{equation}
M_{\rho}^{2} = {{2 {\rho_{d}/\rho_{u}} + \gamma +1} \over {\gamma+1-{\rho_{d}/\rho_{u}} (\gamma-1)}},
\end{equation}
where $\gamma$ is assumed to be 4/3  (Liu et al. 2006; Gopalswamy \& Yashiro 2011).

We calculated the Mach number from both the methods:  (1) Mach
number from the standoff distance ratio, $M_{\Delta R}$ and (2) from
the density compression ratio, $M_{\rho}$. Figure 6 shows the Mach
numbers determined by equations (4) and (5) for the 26 shock
positions in the 10 CMEs. $M_{\Delta R}$ is randomly scattered in
the range of 1.49 and 3.43 with a mean value of 2.41, but $M_{\rho}$
occupies a narrow range of 1.47 and 1.90 with a mean value of 1.56.
We list $M_{\Delta R}$ and $M_{\rho}$ in the ninth and tenth columns
of Table 1, respectively. If we set  $\gamma$ as 5/3, then the
denominators of equation (4) and (5) are close to or below 0 for
very strong shocks ($\Delta R/R_{s} \leq 0.2$ or $\rho_{d}/\rho_{u}
\geq 4$ ), which makes the Mach number unrealistically high.

\section{Results}

\subsection{Alfven Speed}

Since we estimated the upstream Mach numbers from the standoff distance ratio and the density compression ratio, the Alfven speed is easily determined using the simple relation,
\begin{equation}
V_{A} = {V_{SH} \over M}, 
\end{equation}
where $V_{A}$ is the upstream Alfven speed.

Figure 7 shows the distribution of Alfven speeds determined by equation (6) using the standoff distance (filled circles) and the density compression (empty circles) methods. The Alfven speeds from the standoff distance ratio, $V_{A\Delta R}$, are roughly scattered in the range of 259 and 982 km s$^{-1}$ (mean=555 km s$^{-1}$) and from the density compression ratio, the Alfven speeds, $V_{A\rho}$ are in the range of 459 and 1367 km s$^{-1}$ (mean=826 km s$^{-1}$). These values are consistent with the factor of 3 variation in Alfven speed derived from radio quiet and radio loud CMEs (Gopalswamy et al. 2008a, c). As seen in the figure, $V_{A\rho}$ values are much higher than the $V_{A\Delta R}$ values.  We list the Alfven speed $V_{A\Delta R}$ and $V_{A\rho}$ in the eleventh and twelfth column of Table 1, respectively.

\subsection{Magnetic Field Strength}

Since the Alfven speed is defined as
\begin{equation}
V_{A}=2 \times 10^{6}\rho^{-1/2} B(km s^{-1}),
\end{equation}
where the magnetic field strength, B can be determined using
\begin{equation}
B={1 \over 2} \times 10^{-6} V_{A} \rho^{1/2} (G).
\end{equation}
To estimate the magnetic field strength in the upper solar corona, we used the Alfven speeds obtained from the standoff distance ratios and the density compression ratios. The other parameter needed in Equation (8) is the upstream plasma density, which can be obtained in a number of ways. Gopalswamy \& Yashiro (2011) used the density at the nose obtained from the pB images. Gopalswamy et al. (2011) used the plasma density given by the lower-frequency branch in type II band splitting. Since we were not able to get appropriate pB images for all the 10 events, we decided to use a density model. We used the Leblanc, Dulk and Bougeret density model (1998),
\begin{equation}
\rho (r) = 3.3 \times 10^{5} r^{-2} +4.1 \times 10^{6} r^{-4} + 8.0 \times 10^{7} r^{-6}.
\end{equation}

Figure 8 shows the magnetic field strengths in the upstream region through which the 10 fast limb CMEs propagate. The magnetic field strengths in the upper solar corona (3 to 15 $R_{s}$) are distributed from 105 to 6 mG (mean=32 mG) based on the standoff distance ratio. When the density compression ratios are used, B is between 163 and 14 mG (mean=47 mG). The distribution of $B_{\Delta R}$  is consistent with the Dulk and McLean's empirical  model (1978), while the distribution of $B_\rho$ is substantially higher than $B_{\Delta R}$. For comparison we included the magnetic field strengths from previous studies (Patzold et al. 1987; Spangler 2005; Cho et al. 2007; Ingleby et al. 2007; Bemporad \& Mancuso 2010). We also plotted the result from the 2008 March 25 using the standoff distance technique (Gopalswamy \& Yashiro 2011). We list the magnetic field strengths from the standoff distance ratio and the density compression ratio in the thirteenth and fourteenth columns of Table 1, respectively. We note that this kind of shock analysis is adoptable to measurement of magnetic field strength in the solar corona and to interpret CME-driven shock structure.

\subsection{Comparison of Shock Parameters}

As shown in  Figure 8, the magnetic field strengths derived from the
density compression ratio are higher than those from the empirical
model and the standoff distance ratio. We speculate that the density
compression ratio might be underestimated due to contributions from
the background density. That is, it is hard to distinguish the
enhanced electron density from the background electron density
accumulated along the line-of-sight especially in the upper coronal
region. Regarding this argument, Figure 4 shows that the observed
density enhancement decreases as the heliocentric distance
increases. Several authors have attempted to get more accurate
density compression ratio by assuming the shock size $l$ along the
line-of-sight (Ontiveros \& Vourlidas 2009; Eselevich \& Eselevich
2011). It is noted that the standoff distance measurements have no
such weakness. In fact the density compression ratio obtained from
type II burst band-splitting has been shown to agree with the
standoff distance method. Therefore we think the uncertainty in the
compression ratio obtained from white-light observations mainly
comes from the assumption of the line-of-sight depth of the shock.

In order to account for the underestimation of the density compression ratio, we multiplied the compression ratio by a factor of 2. The resulting Alfven speeds are shown in Figure 9. The comparison shows that the Alfven speeds are consistent with each other with a correlation coefficient of 0.74, when the 2-fold density compression ratio is used. Figure 10 shows the comparison between the magnetic field strength from standoff distance ratio and those from the original and 2-fold density compression ratios. The magnetic field strengths from both methods, when the 2-fold density ratio is used, are very consistent with each other with a correlation coefficient of 0.92.

\section{Summary and Conclusion}

To measure the magnetic field strength in the solar corona, we examined 10 fast ($\geq 1000$ km s$^{-1}$) limb CMEs which show clear shock structures in SOHO/LASCO images. By applying the piston-shock relationship to the observed CME's standoff distance, we obtained the coronal Alfven speed to be in the range from 259 to 982 km s$^{-1}$ and the magnetic field strength in the range from 6 to $105 mG$ in the heliocentric distance range of 3 and $15 R_s$. The magnetic field strength are consistent with the empirical models (Dulk \& McLean 1978) and other studies (Patzold et al. 1987; Spangler 2005; Cho et al. 2007; Ingleby et al. 2007; Bemporad \& Mancuso 2010; Gopalswamy \& Yashiro 2011). These results confirm that  the standoff distance ratio provides us with a useful tool to derive the magnetic fields in the wide range of solar corona ($\sim 30 R_s$).

The Alfven speeds and magnetic field strengths derived from the
density compression ratio are about two times higher than the above
results. We speculate that the density compression ratio obtained
from white-light observations might be underestimated since the
observed density is based on the electrons integrated over the
line-of-sight, while the measurements of standoff distance ratio
have no such weakness. To inspect the line-of-sight effect on the
density compression ratio, we adopted the method proposed by
Eselevich \& Eselevich (2011), which uses the differential
brightness dP from the LASCO tB image and the shock size $l$ along
the line-of-sight. Assuming that the CME and Shock's configuration
is symmetric, we measured the length of tangential line at CME’s
nose, which is considered as equivalent to the length along the
line-of-sight, $l$. As the result, we found that if we choose
Newkirk density model (Newkirk 1961) for upstrem density, the
density compression ratio, which is calculated from dP and $l$, has
very good correlation with our previous result from Van de Hulst
inversion method with a correlation coefficient of 0.96 as shown in
Figure 11.

We also note that when we multiply the density compression ratio by
a factor of 2,  the Alfven speeds and the magnetic field strengths
are consistent with those from the standoff distance technique. This
supports the idea that the diffuse structures surrounding the CME
front, as shown in Figure 1, can be interpreted as shock structures
- shock sheath to be precise. There are two main observational
results that support the existence of shocks in the low corona. Type
II radio bursts in the metric (Cliver et al. 1999) and longer
wavelengths (Gopalswamy et al., 2005) are good indicators near the
Sun. The white light observations of diffuse features surrounding
the CME flux ropes confirm this, as inferred from streamer
deflections (Gosling et al. 1974) and other white-light signatures
(Vourlidas et al. 2003; Sheeley et al. 2000; Gopalswamy et al.
2008a; Gopalswamy et al. 2009a) and MHD simulations (Manchester
2009).

Finally, we would like to stress on the fact that this study is a new attempt (together with Gopalswamy \& Yashiro 2011) to measure the magnetic field strengths in the upper corona up to $30 R_s$ by applying the piston-shock relationship to CME coronagraph observations. This method can be applied to CMEs showing clear shock structures surrounding the CME front so that it can provide us a useful method to derive the magnetic fields in the  solar corona.  It is a unique method to derive magnetic fields in the upper solar corona (10 $\sim 20 R_s$).

\acknowledgments

\clearpage

\begin{figure}
\noindent
\includegraphics[width=39pc]{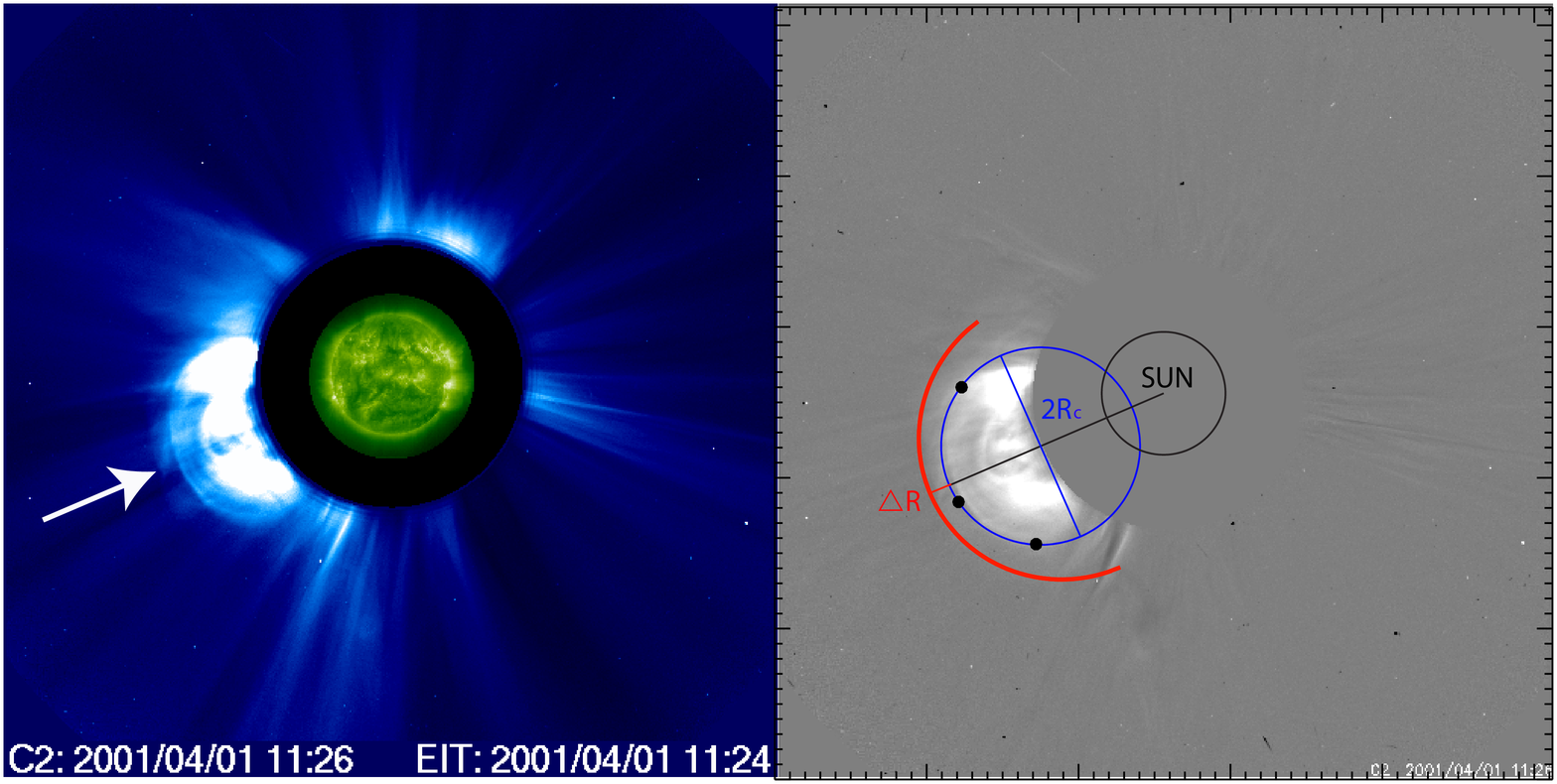}
\caption{The left panel shows a SOHO/LASCO image of the 2001 April 1
CME at 11:26 UT showing the diffuse structure ahead of the CME flux
rope (the sharp feature). The arrow indicates the shock nose. The
right panel shows the running difference image of the event. The
radial black line marks the central position angle (PA) of the shock
nose and the blue circle indicates the CME as an obstacle. The red
lines indicate the shock front and standoff distance,  $\Delta R$.}
\end{figure}

\begin{figure}
\noindent
\includegraphics[width=20pc]{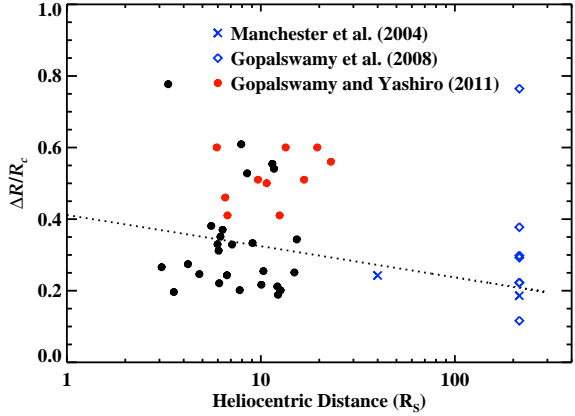}
\caption{The heliocentric distribution of the standoff distance ratio, $\Delta R / R_{c}$, for 26 shock positions of 10 CMEs indicated by black circles. The x-axis is height of the shock position and y-axis is $\Delta R / R_{c}$.  The dotted line presents the polynomial fitting and the red circles indicate the standoff distance ratios for a single event from Gopalswamy \& Yashiro (2011). Diamonds and cross marks represent those from MC observations and the numerical model, respectively.}
\end{figure}

\begin{figure}
\noindent
\includegraphics[width=20pc]{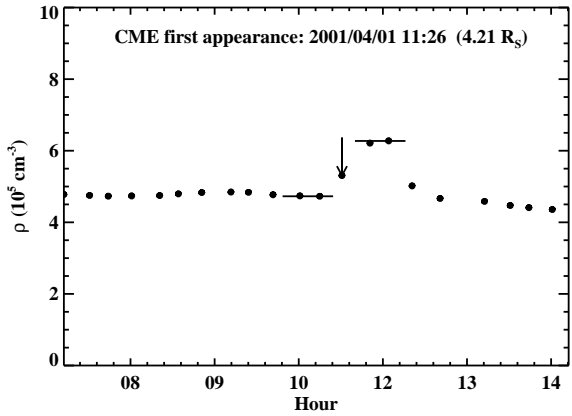}
\caption{An Example of the temporal variation of electron density at the shock position. The arrow indicates the CME's first appearance time and two solid lines show the upstream (left) and downstream (right) electron densities.}
\end{figure}

\begin{figure}
\noindent
\includegraphics[width=20pc]{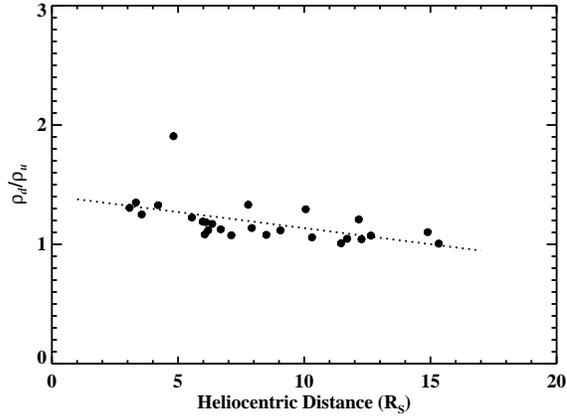}
\caption{The downstream/upstream electron density compression ratio, $\rho_{d}/\rho_{u}$, as a function of the heliocentric distance. The dotted line is the polynomial fitting to the compression ratio.}
\end{figure}

\begin{figure}
\noindent
\includegraphics[width=20pc]{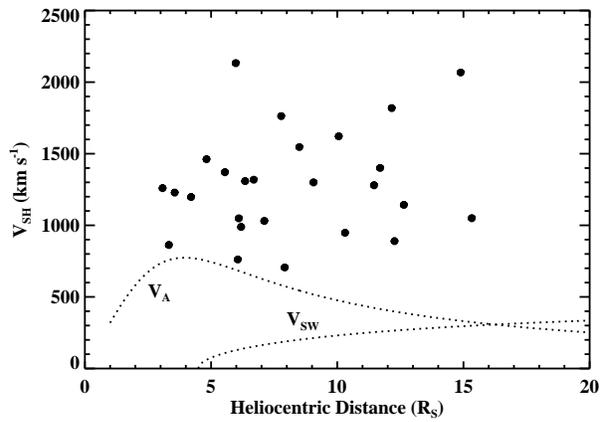}
\caption{The distribution of the shock speeds, $V_{SH}$. The dotted lines show the variations of Alfven speed and the solar wind speed as a function of the heliocentric distance. }
\end{figure}

\begin{figure}
\noindent\includegraphics[width=20pc]{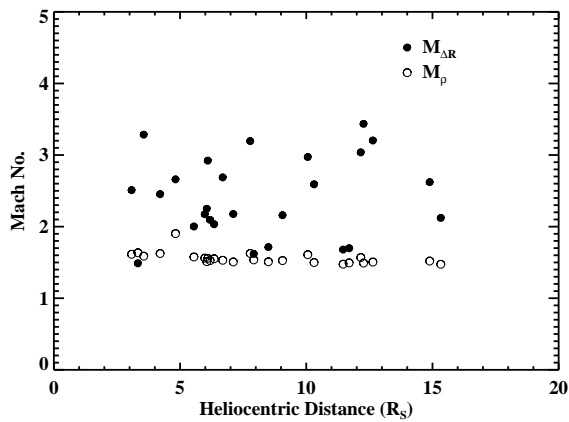}
\caption{The heliocentric distribution of Mach numbers calculated by using the equations (4) and (5). The filled circles indicate the Mach numbers from the standoff distance ratio, $M_{\Delta R}$, and the empty circles indicate those from the density compression ratio, $M_{\rho}$.}
\end{figure}

\begin{figure}
\noindent\includegraphics[width=20pc]{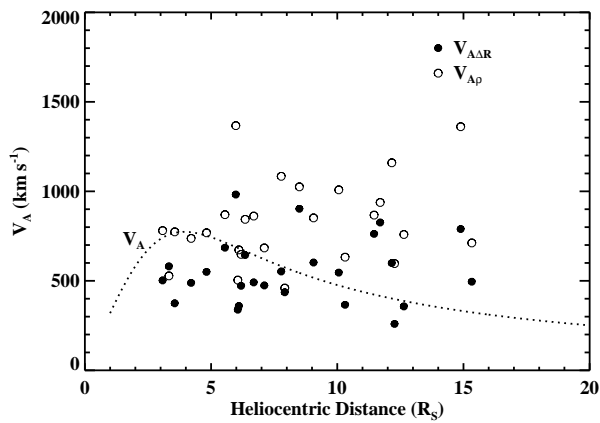}
\caption{The heliocentric distribution of Alfven speeds using the standoff distance ratio, $V_{A\Delta R}$, and the density compression ratio, $V_{A\rho}$. The dotted line indicates the Alfven speed from the model (see Gopalswamy et al. 2001; Eselevich \& Eselevich 2008).}
\end{figure}

\begin{figure}
\noindent\includegraphics[width=20pc]{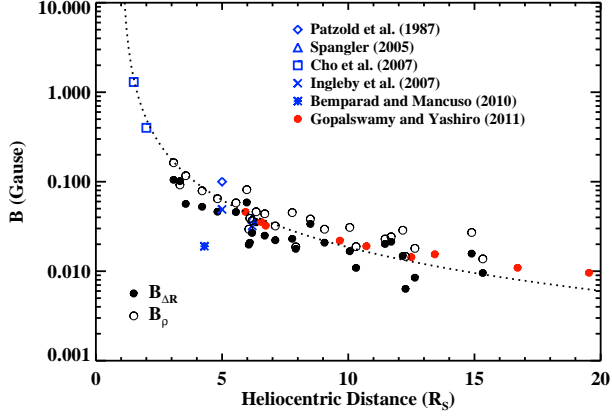}
\caption{Magnetic field strength using the standoff distance ratio, $B_{\Delta R}$, and density compression ratio, $B_\rho$ as a function of heliocentric distance. The dotted line indicates the empirical magnetic field model (Dulk \& McLean 1978).}
\end{figure}

\begin{figure}
\noindent\includegraphics[width=16pc]{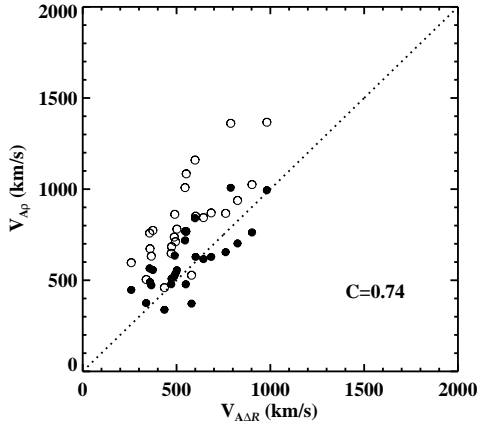}
\caption{The relationship between the Alfven speeds from the standoff distance ratio and the density compress ratios when we take the original and 2-fold density compression ratios. The filled circles indicate the values using 2-fold density compression ratio and the empty circles indicate those from original density compression ratio. The dotted line indicates when Alfven speeds are the same in both methods.}
\end{figure}

\begin{figure}
\noindent\includegraphics[width=16pc]{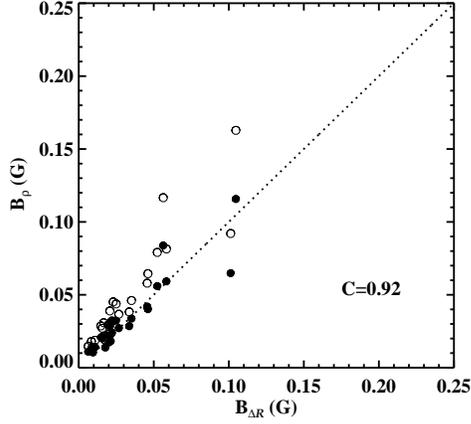}
\caption{The relationship between the magnetic field strengths from the standoff distance ratio and the density compression ratio methods when we take the original and 2-fold density compression ratios. The explanations are the same as Figure 9.}
\end{figure}

\begin{figure}
\noindent\includegraphics[width=16pc]{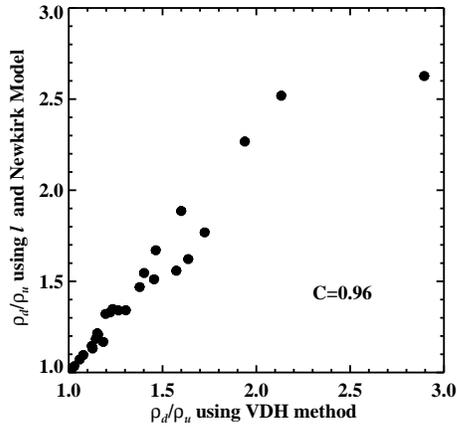}
\caption{The correlation between previous $\rho_d / \rho_u$ values based on VDH inversion method and new $\rho_d / \rho_u$ values based on $l$ and Newkirk density model}
\end{figure}

\clearpage

\begin{deluxetable}{cccccccccccccc}
\tabletypesize{\scriptsize}
\setlength{\tabcolsep}{0.02in}
\tablecaption{Information of very fast ($\ge$ 1000 km s$^{-1}$) limb CMEs which show the clear shock structures from 1997 to 2003.}
\tablewidth{0pt}
\tablehead{
\multicolumn{2}{c}{CME} & \colhead{Type II$^\dagger$} & \multicolumn{2}{c}{Shock Position} & \multicolumn{3}{c}{Shock Parameter} & \multicolumn{2}{c}{Mach No.} & \multicolumn{2}{c}{$V_{A}$ (km s$^{-1}$)} & \multicolumn{2}{c}{$B (mG)$}\\
\colhead{No.} & \colhead{Date/Time} & \colhead{Time} & \colhead{PA ($^\circ$)} & \colhead{Height ($R_s$)} & \colhead{$\Delta R/R_{c}$} & \colhead{$\rho_{d}/\rho_{u}$} & \colhead{$V_{SH}$ (km s$^{-1}$)} & \colhead{$M_{\Delta R}$} & \colhead{$M_{\rho}$ } & \colhead{$V_{A\Delta R}$} & \colhead{$V_{A\rho}$} & \colhead{$B_{\Delta R}$} & \colhead{$B_{\rho}$}}
\startdata
 1 & 1997/11/14 10:14:03 & No & 70 & 3.33 & 0.78 & 1.35 & 862 & 1.49 & 1.63 & 580 & 527 & 101 & 92\\
    & 1997/11/14 10:52:30 &    & 67 & 6.19 & 0.35 & 1.12 & 988 & 2.09 & 1.53 & 471 & 647 &   27 & 37\\
\tableline
 2 & 1999/07/25 13:31:21 & 13:21 M & 301 & 3.08 & 0.27 & 1.30 & 1259 & 2.51 & 1.61 & 501 & 780 & 105 & 163\\
   & 1999/07/25 13:54:05 &            & 301 & 5.55 & 0.38 & 1.22 & 1370 & 2.00 & 1.58 & 684 & 869 & 46 & 58\\
   & 1999/07/25 14:18:05 &            & 301 & 8.50 & 0.53 & 1.08 & 1546 & 1.71 & 1.51 & 902 & 1025 & 34 & 38\\
   & 1999/07/25 14:42:05 &            & 301 & 11.7 & 0.54 & 1.05 & 1400 & 1.70 & 1.49 & 825 & 937 & 21 & 24\\
\tableline
 3 & 2000/04/04 16:43:01 & 15:31 M & 325 & 11.46 & 0.55 & 1.01 & 1279 & 1.68 & 1.48 & 761 & 867 & 20 & 23\\
   & 2000/04/04 17:18:06 &  15:45 D & 326 & 15.33 & 0.34 & 1.00 & 1049 & 2.12 & 1.47 & 494 & 711 & 10 & 14\\
\tableline
 4 & 2000/05/04 11:42:05 & 11:06 M & 226 & 7.92 & 0.61 & 1.14 & 705 & 1.62 & 1.53 & 435& 459 & 18 & 19\\
   & 2000/05/04 12:42:05 & 11:10 D & 226 & 12.64 & 0.20 & 1.07 & 1142 & 3.20 & 1.51 & 356 & 758 & 8 & 18\\
\tableline
 5 & 2000/05/05 16:18:05 & 16:35 D & 213 & 6.35 & 0.37 & 1.17 & 1308 & 2.03 & 1.55 & 643 & 843 & 35 & 46\\
   & 2000/05/05 16:42:06 &     & 214 & 9.06 & 0.33 & 1.12 & 1299 & 2.16 & 1.53 & 601 & 851 & 21 & 29\\
\tableline
 6 & 2000/06/15 20:26:06 & 19:43 M & 306 & 6.06 & 0.31 & 1.08 & 761 & 2.25 & 1.51 & 338 & 504 & 20 & 29\\
   & 2000/06/15 20:42:05 & 19:52 D & 307 & 7.11 & 0.33 & 1.07 & 1030 & 2.17 & 1.51 & 473 & 683 & 22 & 32\\
   & 2000/06/15 21:18:05 &            & 307 & 10.31 & 0.25 & 1.06 & 947 & 2.59 & 1.50 & 365 & 632 & 11 & 19\\
   & 2000/06/15 21:42:05 &            & 308 & 12.27 & 0.19 & 1.04 & 889 & 3.43 & 1.49 & 258 & 596 & 6 & 15\\
\tableline
 7 & 2001/04/01 11:26:06 & No & 116 & 4.21 & 0.27 & 1.33 & 1197 & 2.45 & 1.62 & 488 & 737 & 52 & 79\\
   & 2001/04/01 11:50:07 &     & 117 & 6.69 & 0.24 & 1.12 & 1318 & 2.69 & 1.53 & 490 & 861 & 25 & 44\\
\tableline
 8 & 2001/12/28 20:30:05 & 19:59 M & 151 & 5.98 & 0.33 & 1.19 & 2132 & 2.17 & 1.56 & 981 & 1366 & 58 & 81\\
   & 2001/12/28 21:18:32 & 20:35 D & 152 & 14.89 & 0.25 & 1.10 & 2067 & 2.62 & 1.52 & 788 & 1360 & 16 & 27\\
\tableline
 9 & 2002/01/14 06:05:05 & 06:08 M & 220 & 4.82 & 0.25 & 1.91 & 1461 & 2.66 & 1.90 & 549 & 768 & 46 & 64\\
   & 2002/01/14 06:30:05 & 06:25 D & 220 & 7.78 & 0.20 & 1.33 & 1762 & 3.19 & 1.63 & 551 & 1083 & 23 & 45\\
   & 2002/01/14 06:45:05 &            & 220 & 10.06 & 0.22 & 1.29 & 1621 & 2.97 & 1.61 & 545 & 1008 & 17 & 31\\
   & 2002/01/14 07:00:06 &            & 218 & 12.16 & 0.21 & 1.21 & 1818 & 3.04 & 1.57 & 598 & 1159 & 15 & 29\\
\tableline
10 & 2003/10/24 03:06:06 & No & 124 & 3.56 & 0.20 & 1.25 & 1228 & 3.28 & 1.59 & 373 & 773 & 56 & 117\\
   & 2003/10/24 03:30:05 &      & 124 & 6.10 & 0.22 & 1.18 & 1048 & 2.92 & 1.56 & 358 & 672 & 21 & 39\\
\enddata
\tablecomments{$^\dagger$ M: Metric Type II radio busrt, D: DH Type II radio busrt}
\end{deluxetable}

\end{document}